\documentclass[prd,twocolumn,showpacs,preprintnumbers,amsmath,amssymb,superscriptaddress]{revtex4}
\usepackage{graphicx}
\usepackage{epsfig}
\usepackage{rotating}
\usepackage{dcolumn}
\usepackage{bm}

\newcommand{\lsim}{\raisebox{-0.13cm}{~\shortstack{$<$ \\[-0.07cm] $\sim$}}~} 
\newcommand{\gsim}{\raisebox{-0.13cm}{~\shortstack{$>$ \\[-0.07cm] $\sim$}}~} 
\def\beq{\begin{equation}}   
\def\eeq{\end{equation}}

\def\nn{\nonumber}
\def\bea{\begin{eqnarray}}
\def\eea{\end{eqnarray}}

\begin{document}

\preprint{LPT-ORSAY-09-47, IFIC-09-0702}

\title{Bottomonium spectroscopy with mixing of $\eta_b$ states and a 
light CP-odd Higgs}

\author{Florian Domingo}
\affiliation{Laboratoire de Physique Th\'eorique, Unit\'e mixte de
Recherche -- CNRS -- UMR 8627, Universit\'e de Paris--Sud, F--91405
Orsay, France.}
\author{Ulrich Ellwanger}
\affiliation{Laboratoire de Physique Th\'eorique, Unit\'e mixte de
Recherche -- CNRS -- UMR 8627, Universit\'e de Paris--Sud, F--91405
Orsay, France.}
\author{Miguel-Angel Sanchis-Lozano}
\affiliation{Instituto de F\'{\i}sica Corpuscular (IFIC)
and Departamento de F\'{\i}sica Te\'orica, 
Centro Mixto Universitat de Val\`encia-CSIC, 
Dr. Moliner 50, E-46100 Burjassot, Valencia, Spain}

\begin{abstract} 

The mass of the $\eta_b(1S)$, measured recently by BABAR, is
significantly lower than expected from QCD predictions for the
$\Upsilon(1S)$ -- $\eta_b(1S)$ hyperfine splitting. We
suggest that the observed $\eta_b(1S)$ mass is shifted downwards due to a
mixing with a CP-odd Higgs scalar $A$ with a mass $m_A$ in the range 9.4 --
10.5~GeV compatible with LEP, CLEO and BABAR constraints. We determine the
resulting predictions for the spectrum of the $\eta_b(nS) - A$ system
and the branching ratios into $\tau^+\,\tau^-$ as functions of
$m_A$.

\end{abstract} 
\maketitle

The BABAR collaboration has recently determined the $\eta_b(1S)$ mass
$m_{\eta_b(1S)}$ with an error of only a few MeV in radiative decays
$\Upsilon \to \gamma\,\eta_b$ of excited $\Upsilon$ states and the
observation of peaks in the photon energy spectrum. The result from
$\Upsilon(3S)$ decays is $m_{\eta_b(1S)} = 9388.9^{+3.1}_{-2.3}\
(\mathrm{stat}) \pm 2.7\ (\mathrm{syst})$~MeV \cite{:2008vj}, and the
result from $\Upsilon(2S)$ decays is $m_{\eta_b(1S)} =
9392.9^{+4.6}_{-4.8}\ (\mathrm{stat}) \pm 1.9\ (\mathrm{syst})$~MeV
\cite{:2009pz}. The average gives \cite{:2009pz}
\beq\label{mobs}
m_{\eta_b(1S)} = 9390.9 \pm 3.1\ \mathrm{MeV}\; ,
\eeq
implying a hyperfine splitting $E_{hfs}(1S) = m_{\Upsilon(1S)} - 
m_{\eta_b(1S)}$ of
\beq\label{ehfsexp}
E_{hfs}^{exp}(1S) = 69.9 \pm 3.1\ \mathrm{MeV}\; .
\eeq

This result can be compared to predictions from QCD. Recent results
based on perturbative QCD are in good agreement with each other and
give $E_{hfs}(1S) = 44 \pm 11\ \mathrm{MeV}$ \cite{Recksiegel:2003fm}
and $E_{hfs}(1S) = 39 \pm 14\ \mathrm{MeV}$ \cite{Kniehl:2003ap}
(whereas $E_{hfs}(1S)$ varies over a
wider range in phenomenological models \cite{Brambilla:2004wf}); in
the following we will use tentatively an average value
\beq\label{ehfsqcd}
E_{hfs}^{pQCD}(1S) = 42 \pm 13\ \mathrm{MeV}
\eeq
which is about two standard deviations away from the experimental
result (\ref{ehfsexp}). The most recent result from (unquenched) lattice
QCD is \cite{Gray:2005ur}
\beq\label{ehfslat}
E_{hfs}^{latQCD}(1S) = 61 \pm 14\ \mathrm{MeV}\; ,
\eeq
which is within $1\,\sigma$ of (\ref{ehfsexp}). However, the
hyperfine splitting is quite sensitive to short distances or hard quark
momenta. It has been argued in \cite{Penin:2009wf} that the perturbative
results in \cite{Kniehl:2003ap} can be used for short distance
corrections of the Wilson coefficient of the corresponding spin-flip
operator measured on the lattice. The additional contribution
$\delta^{hard} E_{hfs}(1S)$ to $E_{hfs}^{latQCD}(1S)$ has been estimated
as $\delta^{hard} E_{hfs}(1S) \sim -20$~MeV \cite{Penin:2009wf},
which brings the lattice result (\ref{ehfslat}) in good agreement with
the perturbative result (\ref{ehfsqcd}). Although this conclusion needs
to be checked within perturbation theory with lattice regularization, we
consider it as a support for the perturbative QCD result
(\ref{ehfsqcd}).

Whereas an explanation of the discrepancy between (\ref{ehfsexp}) and
(\ref{ehfsqcd}) within QCD is not excluded at present, we will elaborate
below the consequences of an explanation of this discrepancy due to new
physics in the form of a mixing of the $\eta_b$ states with a CP-odd
Higgs scalar $A$ with a mass $m_A$ in the range $9.4 - 10.5$~GeV: as a
result of such a mixing, the masses of the $\eta_b$~-~like eigenstates
of the full mass matrix can differ considerably from their values in
pure QCD without the presence of the CP-odd Higgs $A$
\cite{Drees:1989du,Fullana:2007uq,Domingo:2008rr}, and the mass of the
state interpreted as the $\eta_b(1S)$ can be smaller than expected if
$m_A$ is somewhat above 9.4~GeV.

In such a scenario, the masses of the states interpreted as $\eta_b(2S)$
and $\eta_b(3S)$ can also be affected, and all states can have
non-negligible branching ratios into $\tau^+\,\tau^-$ due to their
mixing with $A$. According to recent results of BABAR
\cite{Aubert:2009ck}, the corresponding branching ratio of the observed
state is below 8\% at 90\%~confidence level. The branching ratio into
$\mu^+\,\mu^-$ would be smaller by a factor $m_\mu^2/m_\tau^2$, and well
below the present upper limit \cite{Aubert:2009cp}. The investigation of
these phenomena is the purpose of the present paper.

A relatively light CP-odd Higgs scalar can appear, e.\,g., in
non-minimal supersymmetric extensions of the Standard Model (SM) as the NMSSM
(Next-to-Minimal Supersymmetric Standard Model) \cite{Dobrescu:2000jt,
Dobrescu:2000yn, Dermisek:2005ar, Dermisek:2005gg, Dermisek:2006wr,
Domingo:2008rr}. Its mass has to satisfy constraints from LEP, where it
could have been produced in $e^+\,e^- \to Z^* \to Z\,H$ and $H \to A\,A$
(where $H$ is a CP-even Higgs scalar). For $m_A > 10.5$~GeV -- where $A$
would decay dominantly into $b\,\bar{b}$ -- and $m_H < 110$~GeV,
corresponding LEP constraints are quite strong \cite{Schael:2006cr}. For
$2\,m_\tau < m_A < 10.5$~GeV, $A$ would decay dominantly into
$\tau^+\,\tau^-$ and values for $m_H$ down to $\sim 86$~GeV are allowed 
\cite{Schael:2006cr} even if $H$ couples to the $Z$~boson with the
strength of a SM Higgs boson. 

In fact, searches for $e^+\,e^- \to Z^* \to Z\,H$ with $H
\to b\,\bar{b}$ indicate a light excess of events (of $\sim 2.3\,\sigma$
significance) for $m_H \sim 95 - 100$~GeV \cite{Schael:2006cr}, which
could be explained by a reduced branching ratio $BR(H \to b\,\bar{b})
\sim 0.1$ and a dominant branching ratio $BR(H \to A\,A) \sim 0.9$
\cite{Dermisek:2005gg, Dermisek:2006wr} if $A$ decays dominantly into
$\tau^+\,\tau^-$. The possible explanation of this excess of events at
LEP is an additional motivation for a CP-odd Higgs scalar with a mass
below 10.5~GeV. Allowing for $m_H$ somewhat below 100~GeV, such a
scenario would also alleviate the ``little fine tuning problem'' of
supersymmetric extensions of the SM~\cite{Dermisek:2005ar,
Dermisek:2005gg, Dermisek:2006wr}.

Finally CLEO \cite{:2008hs} and BABAR \cite{Aubert:2009ck}
have searched for a light CP-odd scalar $A$
with $A \to \tau^+\,\tau^-$ in radiative $\Upsilon(1S)$ and $\Upsilon(3S)$
decays respectively, which mainly
constrains the range $m_A \lsim 9.4$~GeV. In this work we focus  
on the range $9.4\ \mathrm{GeV} \lsim m_A < 10.5$~GeV, which would be
the most relevant for strong $\eta_b - A$ mixing effects as advocated 
in~\cite{Domingo:2008rr}.

Apart from $m_A$, such mixing effects depend in a calculable way on the
model-dependent coupling of $A$ to $b$~quarks. The corresponding
coupling, normalized with respect to the coupling of the SM
Higgs scalar to $b$~quarks, will be denoted by $X_d$. In models with two
Higgs doublets $H_u$ and $H_d$, where $H_u$ couples to up~quarks and
$H_d$ to down~quarks and leptons, one has 
\cite{Dobrescu:2000jt,Dobrescu:2000yn, Dermisek:2005ar, Dermisek:2005gg, Dermisek:2006wr,Domingo:2008rr}
\beq\label{xd}
X_d = \cos\theta_A\ \tan\beta
\eeq
where $\cos\theta_A$ denotes the SU(2)~doublet component of the CP-odd
scalar $A$, and $\tan\beta = \left<H_u\right>/\left<H_d\right>$. For
$\tan\beta \gg 1$, $X_d$ can equally satisfy $X_d \gg 1$
\cite{Domingo:2008rr}.

Below we proceed as follows: i) First we assume that, in the absence of
a CP-odd Higgs scalar, $m_{\eta_b^0(1S)}$ would have a value
compatible with (\ref{ehfsqcd}), i.\,e.
\beq\label{metaqcd}
m_{\eta_b^0(1S)} \sim 9418 \pm 13\ \mathrm{MeV}\; .
\eeq
ii) We diagonalise the $\eta_b(nS)-A$ mass matrix ($n=1,2,3$) and
require that one eigenvalue coincides with the mass measured by BABAR
within errors (\ref{mobs}); this condition gives us an allowed strip in
the $X_d - m_A$ plane, which depends only weakly on the assumed masses
of $\eta_b^{0}(2S)$ and $\eta_b^{0}(3S)$. 

The resulting values of $X_d$ as function of $m_A$ allow to
determine the remaining eigenvalues of the mass matrix, and the
decompositions of the eigenvectors in terms of $A$ and ${\eta_b^0}(nS)$.
Finally the $A$~components of the eigenstates allow us to determine
their partial widths and to estimate the branching ratios into
$\tau^+\,\tau^-$.

For the $\eta_b^0(1S) - \eta_b^0(2S) - \eta_b^0(3S) - A$ mass matrix we
make the ansatz \beq\label{eq:massmatr}
{\cal M}^2=  
\left(
     \begin{array}{cccc}
     m_{\eta_b^0(1S)}^2 & 0 & 0 & \delta m_1^2\\
     0 & m_{\eta_b^0(2S)}^2 & 0 &\delta m_2^2\\
     0 & 0 & m_{\eta_b^0(3S)}^2 & \delta m_3^2\\
     \delta m_1^2 & \delta m_2^2 & \delta m_3^2 & m_A^2
     \end{array}
\right)\; .
\end{equation}
The diagonal elements $m_{\eta_b^0(nS)}^{2}$ are assumed to be known
(within errors) from QCD with (\ref{metaqcd}) for  $m_{\eta_b^0(1S)}$.
The results depend only weakly on the $\eta_b^{0}(2S)$ and 
$\eta_b^{0}(3S)$ masses, for which we take \cite{Recksiegel:2003fm}
$m_{\eta_b^0(2S)} = 10002$~MeV, $m_{\eta_b^0(3S)} = 10343$~MeV. (We
neglect finite width effects in ${\cal M}^2$, and take real matrix
elements.)

The off-diagonal elements $\delta m_n^2$ can be compu\-ted in the
framework of a non-relativistic quark potential model in terms of the
radial wave functions at the origin \cite{Drees:1989du,Fullana:2007uq}.
These can be considered as identical for vector and pseudocalar states,
and be determined from the measured $\Upsilon \to e^+e^-$
decay widths. Substituting recent values for these widths (see
\cite{Domingo:2008rr} for details) we find
\bea
 &\delta m_1^2&\ \simeq\ (0.14\pm 10\%)\ \mathrm{GeV}^2\times X_d\;,\nn \\
 &\delta m_2^2&\ \simeq\ (0.11\pm 10\%)\ \mathrm{GeV}^2\times X_d\;,\nn \\
 &\delta m_3^2&\ \simeq\ (0.10\pm 10\%)\ \mathrm{GeV}^2\times X_d\;.
 \label{dmest}
\eea
We estimated the errors from higher order QCD corrections to the
relation between the radial wave functions at the origin and
$\Gamma(\Upsilon(nS) \to e^+e^-)$ \cite{Drees:1989du,Fullana:2007uq} to
be $\sim 10\%$. (These errors play only a minor role
for our results.)

\begin{figure}[ht!]
\begin{center}
\includegraphics*[width=0.6\linewidth,height=0.6\linewidth]
{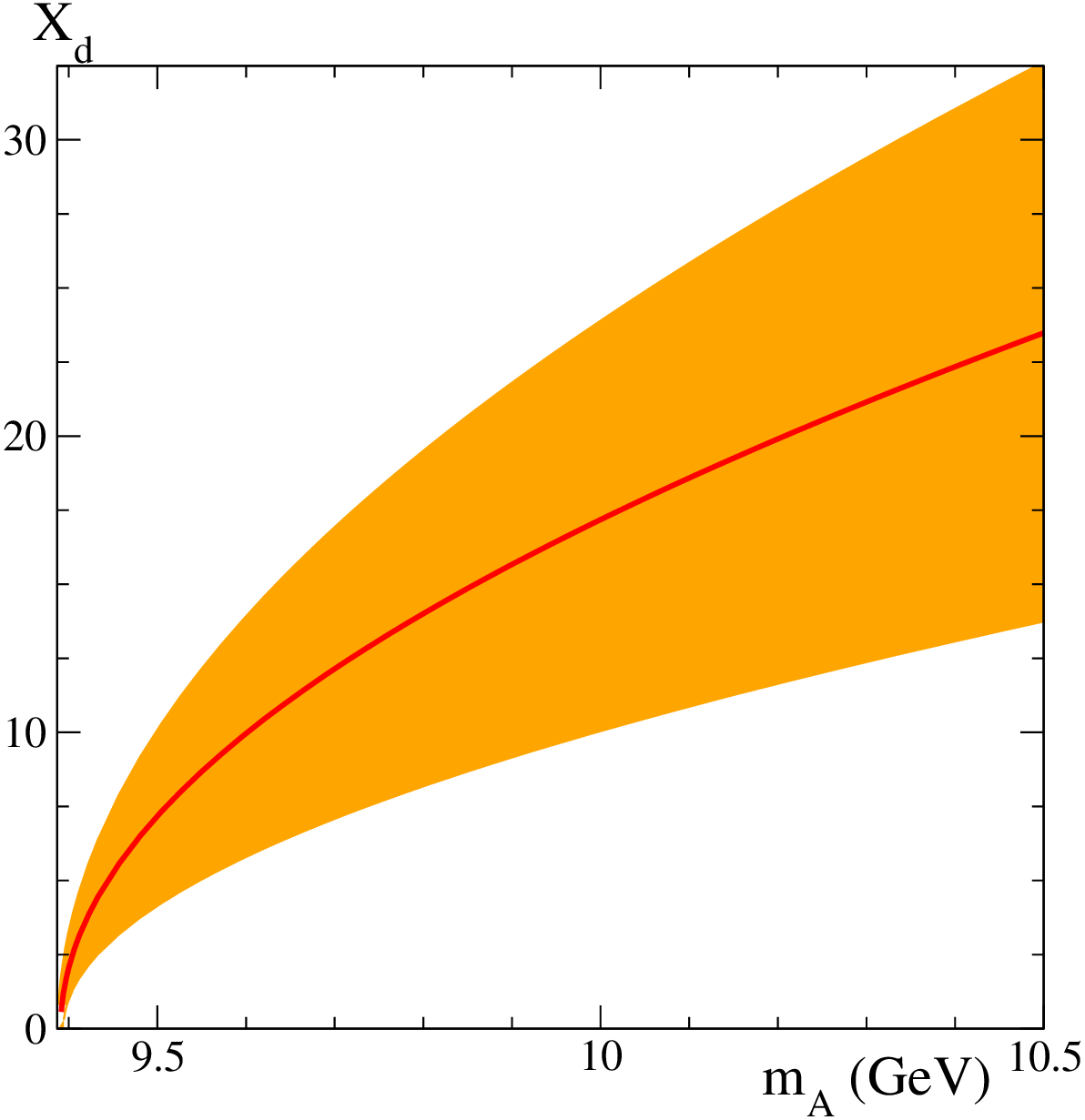}
\end{center}
\addvspace{-6mm}
\caption{$X_d$ as a function of $m_A$ (in GeV) such that one eigenvalue
of ${\cal M}^2$ coincides with the BABAR result (\ref{mobs}).}
\label{fig:xd_ma}
\end{figure}

In order that one eigenvalue of ${\cal M}^2$ coincides with the BABAR
result (\ref{mobs}) subsequently denoted as $m_{obs}$,  $m_A$ in
(\ref{eq:massmatr}) has to satisfy
\bea
m_A^2 &=& m_{obs}^2 
+ \frac{\delta m_1^4}{m_{\eta_b^0(1S)}^2 - m_{obs}^2}\nn \\
&+& \frac{\delta m_2^4}{m_{\eta_b^0(2S)}^2 - m_{obs}^2}
+ \frac{\delta m_3^4}{m_{\eta_b^0(3S)}^2 - m_{obs}^2}\; .
\label{eq:ma}
\eea
Once $m_A$ is expressed in terms of $X_d$,
$X_d$ remains the only unknown parameter in  ${\cal M}^2$.
Varying $m_{obs}$ within the errors in (\ref{mobs}),
$m_{\eta_b^0(1S)}$ within the errors in (\ref{metaqcd}) and 
$\delta m_i^2$ within the errors in (\ref{dmest}), we obtain
for $X_d$ as a function of $m_A$ the result shown in 
Fig.~\ref{fig:xd_ma}. (Recall that $m_A$ -- or, the heaviest eigenstate
of ${\cal M}^2$ with a large $A$~component -- has to be below 10.5~GeV
in order to satisfy LEP constraints in the presence of a Higgs scalar
with a mass around 100~GeV.)

\begin{figure}[ht!]
\begin{center}
\includegraphics*[width=0.75\linewidth]
{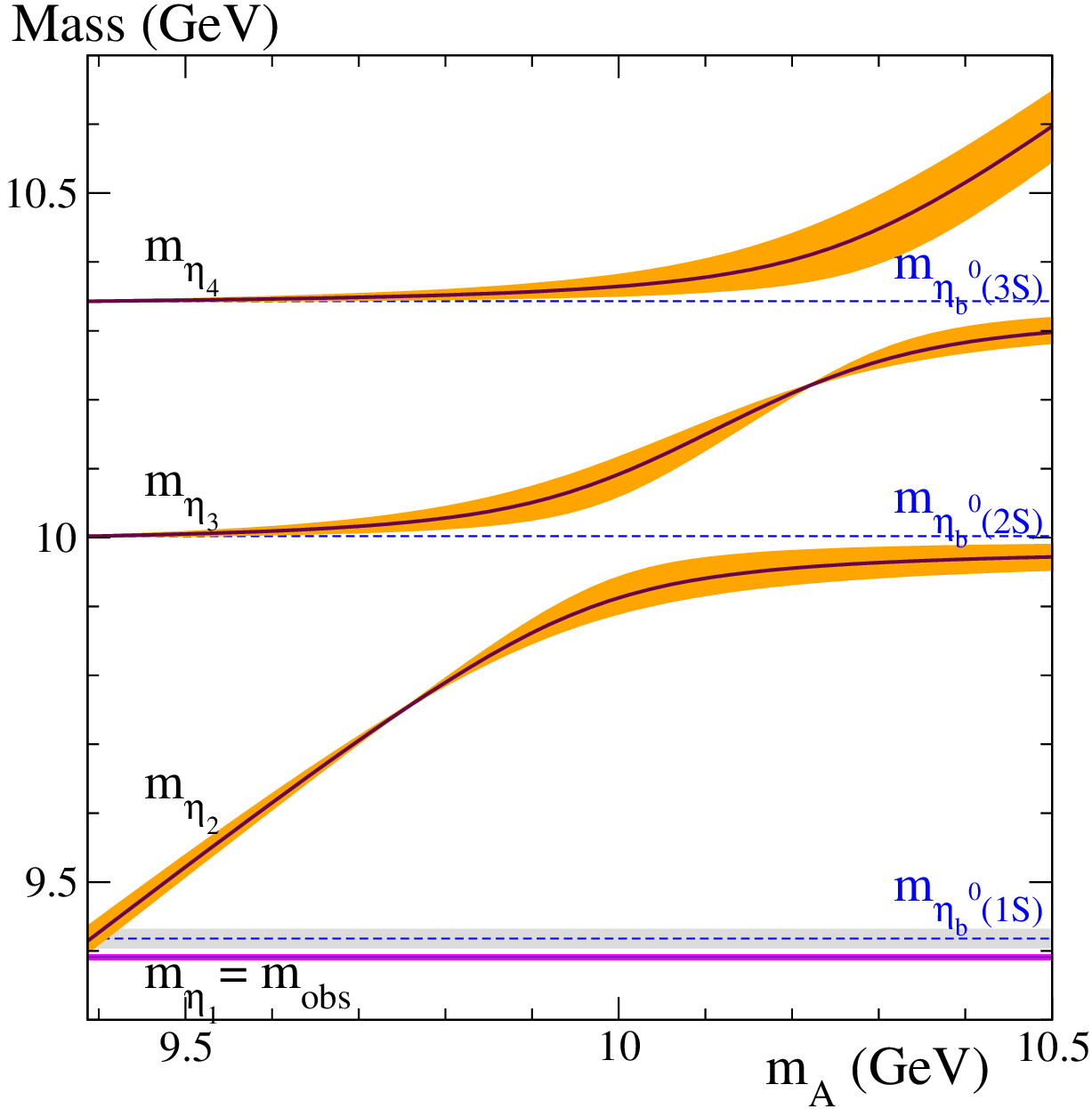}
\end{center}
\addvspace{-6mm}
\caption{The masses of all eigenstates as function of $m_A$.}
\label{fig:heavy_masses}
\end{figure}

Now the masses of all 4 eigenstates of ${\cal M}^2$ can be computed,
which are shown together with the error bands from $m_{obs}$,
$m_{\eta_b^0(1S)}$ and $\delta m_i^2$ (in orange/grey) in
Fig.~\ref{fig:heavy_masses} as functions of $m_A$. Henceforth we
denote the 4 eigenstates of ${\cal M}^2$ by $\eta_i$, $i=1\dots 4$
where, by construction, $m_{\eta_1} \equiv m_{obs}$. For clarity we have
indicated in Fig.~\ref{fig:heavy_masses} our assumed values for
$m_{\eta_b^0(nS)}$ as horizontal dashed lines. For $m_A$ not far above
9.4~GeV (where $X_d$ is relatively small) the effects of the mixing on
the states $\eta_b^0(2S)$ and $\eta_b^0(3S)$ are negligible, but for
larger $m_A$ the spectrum can differ considerably from the one expected
without the presence of $A$.

Now we consider the branching ratios of the eigenstates into
$\tau^+\,\tau^-$, which are induced by their $A$-components.
The decomposition of the eigenstates into the states before mixing can
be written as
\beq
\eta_i = P_{i,1}\;\eta_b^0(1S)
+ P_{i,2}\;\eta_b^0(2S)+ P_{i,3}\;\eta_b^0(3S) + P_{i,4}\;A\;.
\eeq
It turns out that the coefficients $P_{i,j}$ can be expressed
analytically in terms of the eigenvalues $m^2_{\eta_i}$ of ${\cal M}^2$:
\bea\label{eq:pij}
P_{i,4} &=& \Bigg[1 
+ \frac{\delta m_1^4}{(m^2_{\eta_b^0(1S)}-m^2_{\eta_i})^2}
\nn \\
&+& \frac{\delta m_2^4}{(m^2_{\eta_b^0(2S)}-m^2_{\eta_i})^2}
+ \frac{\delta m_3^4}{(m^2_{\eta_b^0(3S)}-m^2_{\eta_i})^2}
\Bigg]^{-1/2}\; ;\nn \\
P_{i,j} &=& 
\frac{-\delta m_j^2}{m^2_{\eta_b^0(jS)}-m^2_{\eta_i}} P_{i,4}
\quad \mathrm{for}\ j=1,2,3\; .
\eea

\begin{figure}[ht!]
\begin{center}
\includegraphics*[width=0.8\linewidth]{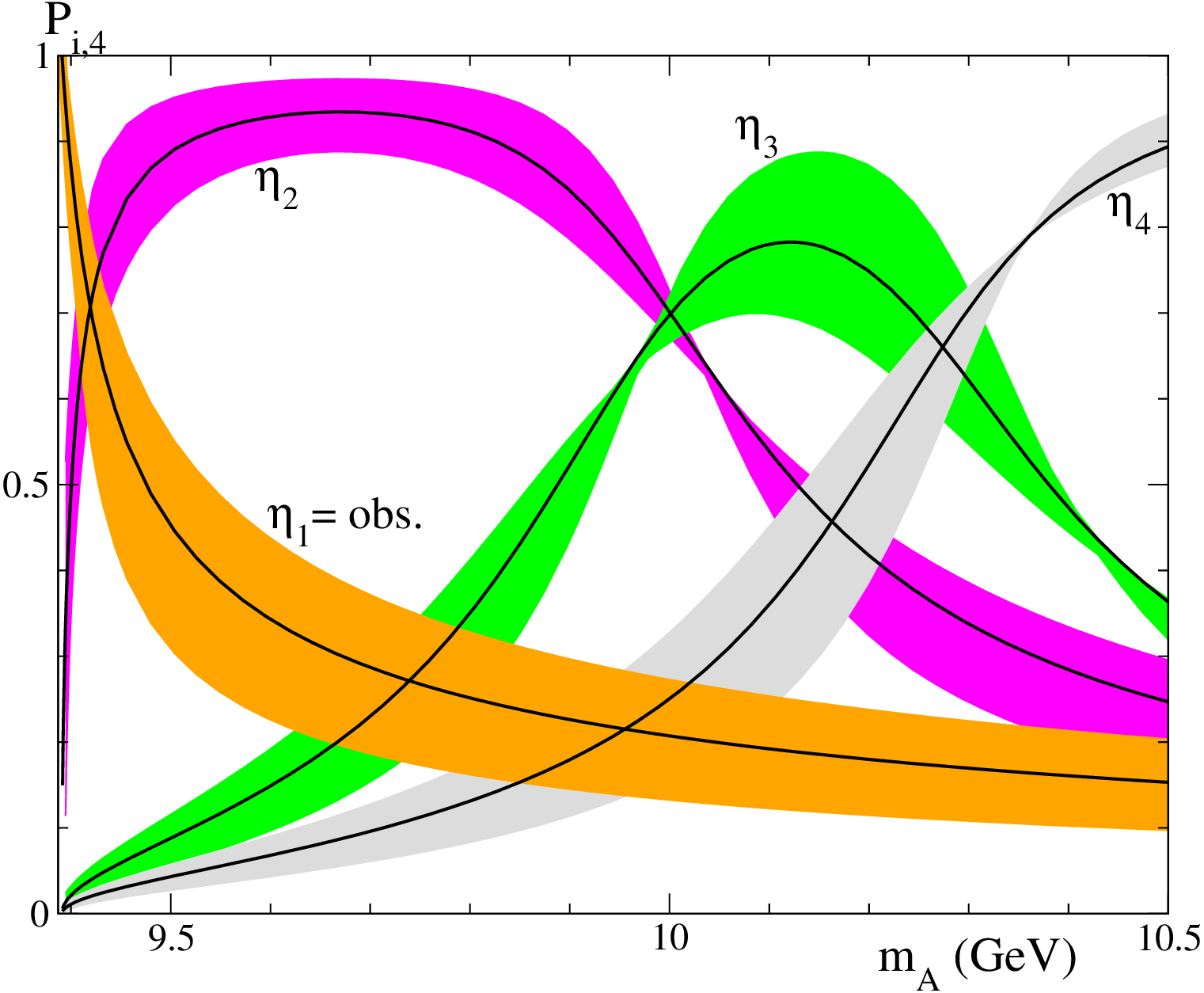}
\end{center}
\addvspace{-6mm}
\caption{The $A$-components
$|P_{i,4}|$ for all 4 eigenstates as functions of $m_A$.}
\label{fig:PAmA}
\end{figure}

In Fig.~\ref{fig:PAmA} we show our results for the $A$-components
$P_{i,4}$ for all 4 eigenstates together with the error bands from
$m_{obs}$, $m_{\eta_b^0(1S)}$ and $\delta m_i^2$.

In the case of $\eta_1 \equiv \eta_{obs}$, only the coefficients
$P_{1,1}$ and $P_{1,4}$ differ significantly from 0. This allows to
express the branching ratio $BR(\eta_1 \to \tau^+\,\tau^-)$ as
\beq\label{eq:br1}
BR(\eta_1 \to \tau^+\,\tau^-) = \frac{P_{1,4}^2 \Gamma_A^{\tau\tau}}
{P_{1,1}^2 \Gamma_{\eta_b^0(1S)} + P_{1,4}^2 \Gamma_A^{tot}}
\eeq
where $\Gamma_A^{\tau\tau}$ is the partial width for $A \to
\tau^+\,\tau^-$, $\Gamma_{\eta_b^0(1S)}$ the width of the state
$\eta_b^0(1S)$ without mixing with $A$, and $\Gamma_A^{tot}$ the total
width of $A$ (without mixing). For $\Gamma_A^{\tau\tau}$ we have
\beq\label{eq:ga}
\Gamma_A^{\tau\tau} = X_d^2 \frac{G_F m_\tau^2 M}{4\sqrt{2}\pi}
\sqrt{1-4\frac{m_\tau^2}{M^2}}
\eeq
where we would have to identify $M$ with $m_A$ if $A$ decays on its
mass shell. Since the dependence on $M$ originates from phase space
integrals, $M$ has to be identified with the mass of the decaying state
(which is actually always close to $m_A$) in our case. Since the $BR(A
\to \tau^+\,\tau^-)$ is typically $\sim  90\%$ (for large $\tan\beta$),
we take $\Gamma_A^{tot} \sim 1.1\; \Gamma_A^{\tau\tau}$ in
(\ref{eq:br1}).

It is remarkable that, once we insert eqs.~(\ref{eq:pij}) and
(\ref{eq:ga}) into the expression (\ref{eq:br1}) for
$BR(\eta_1 \to \tau^+\,\tau^-)$, all dependence on $X_d$ cancels: even
for $m_A \to 9.4$~GeV where $P_{1,4}^2 \to 1$, one has $\Gamma_A^{tot}
\to 0$ from $X_d \to 0$, and hence the first term in the denominator in
(\ref{eq:br1}) always dominates. The numerical result is
\beq\label{eq:br1res}
BR(\eta_1 \to \tau^+\,\tau^-) = (2.4^{+2.3}_{-2.0})\times 10^{-2} \times
\left(\frac{10\,\mathrm{MeV}}{\Gamma_{\eta_b^0(1S)}}\right)\; .
\eeq
Since the analytic expression for this $BR$ is proportional to 
$(m^2_{obs}-m^2_{\eta_b^0(1S)})^2$ in our approach, its lowest possible
value is quite sensitive to the smallest allowed value for this
difference.

Using $\Gamma_{\eta_b^0}(nS)/\Gamma_{\eta_c}(nS)
\simeq  (m_b/m_c)[\alpha_s(m_b)/\alpha_s(m_c)]^5$ $
\simeq 0.25-0.75$~\cite{oliver} and $\Gamma_{\eta_c}(1S) = 26.7\pm 3$~MeV
\cite{Amsler:2008zz} we estimate $\Gamma_{\eta_b^0(1S)} \sim 5-20$~MeV.
Hence the predicted branching ratio is compatible with BABAR upper
limit of 8\% \cite{Aubert:2009ck}.

\begin{figure}[ht]
\begin{center}
\includegraphics*[width=0.8\linewidth]{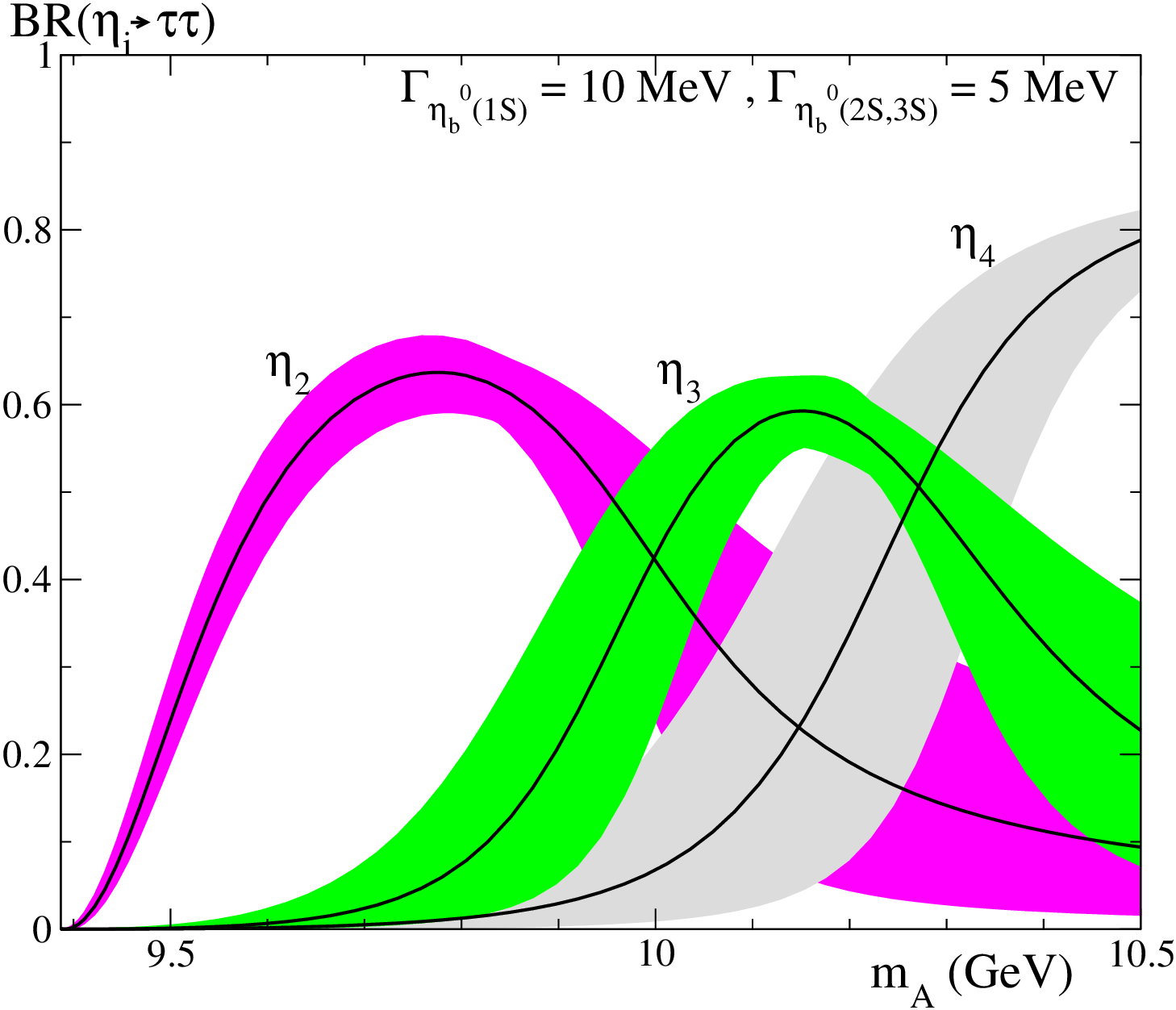}
\end{center}
\addvspace{-6mm}
\caption{The branching ratios into $\tau^+\,\tau^-$ for the eigenstates
$\eta_2$, $\eta_3$ and $\eta_4$ as functions of $m_A$.}
\label{fig:BrmA}
\end{figure}

Turning to the remaining heavier eigenstates, expressions similar to
(\ref{eq:br1}) are always very good approximations for the branching
ratios into $\tau^+\,\tau^-$, since the eigenstates consist essentially
of just one $\eta_b^0(nS)$ state and the CP-odd Higgs~$A$. However, the
branching ratios vary now with $m_{\eta_i}$ and hence with $m_A$;
the results are shown graphically in Fig.~\ref{fig:BrmA} assuming
$\Gamma_{\eta_b^0(1S)} \sim 10$~MeV and 
$\Gamma_{\eta_b^0(2S)} \sim \Gamma_{\eta_b^0(3S)} \sim 5$~MeV. For
larger (smaller) total widths these branching ratios would be somewhat
smaller (larger).

An important issue is the production rate of the eigenstates $\eta_i$ in
radiative decays of excited $\Upsilon$ states, notably in $\Upsilon(3S)
\to \gamma\,\eta_i$. For a pure CP-odd scalar $A$,
the ${BR} \left(\Upsilon(3S) \to \gamma\,
A\right)$ is given by the Wilczek formula~\cite{Wilczek:1977pj}
\beq\label{eq:wilczek}
\frac{{BR} \left(\Upsilon(nS) \to \gamma\, A\right)}
{{BR} \left(\Upsilon(nS) \to \mu^+\mu^-\right)} =
\frac{G_F m_b^2 X_d^2}
{\sqrt{2}\pi\alpha}\biggl(1-\frac{m_{A}^2}{m_{\Upsilon(3S)}^2}\biggr)
\times  F\; .
\eeq
$F$ is an
$m_A$ dependent correction factor, which includes three kinds of
corrections (the relevant formulas
are summarized in \cite{guide}): bound state, QCD, and relativistic
corrections. Unfortunately these corrections become unreliable for $m_A
\gsim 8$~GeV; a na\"ive extrapolation of the known corrections leads to a
vanishing correction factor $F$ for $m_A \gsim 8.8$~GeV 
\cite{Domingo:2008rr} as relevant here.

Thus it is difficult to predict the branching ratios ${BR}
\left(\Upsilon(3S) \to \gamma\, \eta_i\right)$: if the ${BR}
\left(\Upsilon(3S) \to \gamma\, A\right)$ is assumed to vanish, the
production of the states $\eta_i$ has to rely on their $\eta_b^0(nS)$
components. Otherwise, negative interference effects could appear
leading to suppressed branching ratios.
Hence it cannot be guaranteed that the (kinematically accessible) part
of the spectrum shown in Fig.~\ref{fig:heavy_masses} is actually visible
in radiative $\Upsilon(3S)$ decays in the form of a peak in the photon
energy spectrum \cite{Aubert:2009ck}.

In view of the possibly quite low photon energies and/or large
backgrounds, the photons can well escape undetected even if the process
$\Upsilon(3S) \to \gamma\, \eta_i$ occurs with a non-negligible rate. In
this case, as advocated in \cite{SanchisLozano:2002pm}, the
$A$~components of $\eta_i$ can still manifest themselves through a
breakdown of lepton universality in the form of an excess of
$\tau^+\,\tau^-$ final states in $\Upsilon(3S) \to l^+\,l^-$
\cite{Fullana:2007uq,Domingo:2008rr}. However, in the case of a
vanishing correction factor $F$ in (\ref{eq:wilczek}), this phenomenon
would disappear as well.

On the other hand, if the existence of a CP-odd scalar $A$ and a CP-even
scalar $H$ with $m_H \sim 95 - 100$~GeV (and a dominant $H \to A\,A$
branching ratio of $\sim 90\%$) is responsible for the excess of events
at LEP as noted above, it becomes important to test this scenario at the
LHC: The standard search channels for a SM-like CP-even
scalar $H$ would
fail, and the final states from $H \to A\,A$ with $m_A$ below the
$b\,\bar{b}$ threshold would be difficult to detect. Proposals for a
verification of this scenario at the LHC have been made recently in
\cite{Forshaw:2007ra, Belyaev:2008gj, Lisanti:2009uy}.

To conclude, if the mixing with a CP-odd Higgs scalar is responsible for
the discrepancy between the BABAR measurement of $m_{\eta_b(1S)}$ and
the expectations from QCD, it can manifest itself in the form of a
completely distorted spectrum of states as shown in
Fig.~\ref{fig:heavy_masses}. The branching ratios  into $\tau^+\,\tau^-$
would be non-vanishing, albeit below the present experimental upper
limit (for the lowest lying state). These manifestations of a light
CP-odd Higgs scalar in $\Upsilon$ physics at (Super) B factories
\cite{Bona:2007qt} would be complementary to its possible discovery at
the LHC.

\vspace{3mm}

\centerline{\bf Acknowledgments}

We gratefully acknowledge discussions with Y.~Kolo\-mensky, S.~Sekula
and A.~Snyder of the BABAR collaboration.
This work was supported in part by the research grant FPA2008-02878.

\end{document}